\documentclass[sn-apa]{sn-jnl}

\usepackage{url}
\usepackage{amsmath}
\usepackage{graphicx}
\usepackage{booktabs}
\usepackage{tabularx}
\usepackage{booktabs}
\usepackage{balance}
\jyear{2023}
\usepackage{fancyhdr} % for arxiv
\fancypagestyle{firststyle}
{
   \fancyhf{}
   \fancyfoot[C]{\scriptsize{\textit{Digital Society}, vol.~2, no.~1, pp.~1--10. This is the authors' copy. The publisher's copy is available online via \url{https://doi.org/10.1007/s44206-023-00041-7}}}
}

\begin{document}

\title[Reflections on the Data Governance Act]{Reflections on the Data Governance Act}

\author*[1]{\fnm{Jukka} \sur{Ruohonen}}\email{juanruo@utu.fi}
\author[2]{\fnm{Sini} \sur{Mickelsson}}\email{sini.mickelsson@utu.fi}
\affil*[1]{\orgdiv{Faculty of Technology}, \orgname{University of Turku},
\orgaddress{\street{Turun yliopisto}, \city{Turku}, \postcode{FI-20014}, \country{Finland}}}
\affil[2]{\orgdiv{Faculty of Law}, \orgname{University of Turku},
\orgaddress{\street{Turun yliopisto}, \city{Turku}, \postcode{FI-20014}, \country{Finland}}}

\abstract{The European Union (EU) has been pursuing a new strategy under the
  umbrella label of digital sovereignty. Data is an important element in this
  strategy. To this end, a specific Data Governance Act was enacted in
  2022. This new regulation builds upon two ideas: reuse of data held by public
  sector bodies and voluntary sharing of data under the label of data
  altruism. This short commentary reviews the main content of the new
  regulation. Based on the review, a few points are also raised about potential
  challenges.}


\maketitle

\section*{Introduction}

\thispagestyle{firststyle} % for arxiv

The Data Governance Act (DGA) is a part of the larger regulatory framework
pursued by the EU for digitalization, data economy, artificial intelligence, and
other important policy goals often approached under the label of digital
sovereignty.\footnote{~It should be mentioned that digital sovereignty is a
  vague and theoretically problematic concept that covers numerous different
  policy domains \citep{Lambach22, Ruohonen21MIND}. \citet{Sheikh22} provides a
  recent attempt to make sense of the concept in the EU context.} Data is a key
ingredient in this framework. Artificial intelligence needs data. Science needs
data. Digital applications and services need data.

Data is the new oil, so the saying goes. But, in many ways, the new economy and
the giant companies built around data share the same notoriety as the ruthless
oil barons of the past~\citep{Lahtiranta18}. Against this backdrop, it is
understandable that the EU's focus has long been on data protection. This focus
culminated to the enactment of the General Data Protection Regulation (GDPR) in
2016. In recent years the focus has switched toward facilitating data economy
and data sharing in Europe. To some extent, however, the data economy focus was
already present during the policy-making of the GDPR. For many politicians and
stakeholders, the regulation had the twin goals of protecting personal data and
facilitating the free flow of personal data across the internal
market~\citep{Konig22}. Data reuse was also a hot topic during the
negotiations~\citep{Starkbaum19}. Another point is that to a certain degree the
data economy focus was also explicitly embedded to the GDPR, which, according to
Article~20, gave data subjects a new right for data portability between
different data controllers. While the idea was to facilitate data sharing and
interoperability, the new portability right turned out to be problematic in many
ways, particularly with respect to data reuse~\citep{vanOoijen19}. To this end,
the DGA seeks to facilitate further sharing of personal data by introducing a
concept of data altruism. Another core tenet in the new regulation is the reuse
of data held by public sector bodies.

The goals of the DGA are ambitious. The primary goal is to facilitate data
economy in Europe and improve the EU's digital single market. Particular
emphasis is placed upon small- and medium-sized enterprises (SMEs) and start-ups
for which the planned data reuse and data sharing provide new material to
innovate in artificial intelligence and digital applications. Scientific
research is also an important part of the goals. In general, data is seen as
necessary for tackling the climate change and facilitating the green transition,
improving the energy infrastructure, healthcare, and financial services, and so
on and so forth. These goals are framed with a distinct ``European way'' to data
and data economy. Therefore, fairness, data protection, and lawfulness receive a
considerable attention in the regulation. In what follows, the main content of
the new regulation is briefly reviewed. After the review, a few reflections are
provided about potential challenges ahead.

\section*{The DGA in Brief}

The Data Governance Act was proposed by the Commission in 2020 and it was
approved by the Parliament in 2022. This Regulation (EU) 2022/868 will apply
from September 2023 \citep[see][]{EC22a}. The first Article in the regulation
specifies the scope. Accordingly, the regulation lays down (1)~the conditions
for reuse of data held by European public sector bodies; (2)~a notification and
supervisory framework for the provision of data intermediation services; (3)~a
framework for voluntary registration mechanism for entities that collect and
process data made available on altruistic purposes; and (4) a framework for
establishing a new European board for innovation in data economy. Data altruism
is defined in the second article; it refers to ``the voluntary sharing of data
on the basis of the consent of data subjects to process personal data pertaining
to them, or permissions of data holders to allow the use of their non-personal
data without seeking or receiving a reward that goes beyond compensation related
to the costs that they incur where they make their data available for objectives
of general interest as provided for in national law, where applicable, such as
healthcare, combating climate change, improving mobility, facilitating the
development, production and dissemination of official statistics, improving the
provision of public services, public policy making or scientific research
purposes in the general interest''. In other words, data altruism is based
either on the permission given by an organization for not-for-profit processing
activities of non-personal data or the notion of consent in case personal data
is involved.

The categories of data for reuse are defined in Article~3. Accordingly, the
regulation applies to data held by public sector bodies which is protected on
the grounds of commercial confidentiality, statistical confidentiality,
protection of intellectual property, and protection of personal data. Thus,
personal data held by public sector bodies is covered and hence also the GDPR
applies. There are also exclusions. The regulation does not cover data held by
public undertakings, data held by public service broadcasters and their
subsidiaries, data held by cultural establishments and educational institutions,
data protected on the grounds of national security and defense, and data falling
outside the scope of the public tasks of the public sector bodies concerned.

The conditions for data reuse are defined in Article 5. The general principles
are non-discrimination, transparency, proportionality, and proper justification
without attempts to restrict competition. To ensure that data is properly
protected, public sector bodies must ensure that personal data is anonymized and
commercially confidential data is properly modified, aggregated or otherwise
handled with proper disclosure controls. Thus, the GDPR's concept of
pseudonymisation is not sufficient: proper anonymization is generally required
for reuse of personal data.\footnote{~Though, interestingly enough, Articles~7
  and 12 still mention pseudonymisation. Also recital 15 notes that reuse of
  pseudonymised data can be considered, provided that re-identification of data
  subjects remains impossible} That said, the fifth article provides also two
alternative options: a secure processing environment controlled by a public
sector body in case remote access is provided or reuse and processing at the
physical premises of a public sector body. In all cases security must be
guaranteed. The fifth article also prohibits users of reused data from any
attempts to re-identify data subjects. To help public sector bodies with their
new tasks, Article~7 specifies that the member states are obligated to designate
specific competent bodies. The support provided by these competent bodies
includes technical guidance for data storage and data processing, help with
anonymization, suppression, randomization, and other techniques that ensure
privacy, confidentiality, integrity, and accessibility of personal data,
state-of-the-art privacy-preserving methods, deletion of commercially
confidential information, support for consent and permission requests for reuse,
and relevant contractual commitments. According to Article~6, public sector
bodies may also charge fees for allowing reuse of data they possess.

The regulation also introduces a concept of data intermediation
services. According to Article~2, a data intermediation service is ``a service
which aims to establish commercial relationships for the purposes of data
sharing between an undetermined number of data subjects and data holders on the
one hand and data users on the other, through technical, legal or other
means''. The keyword is commercial relationships; other data sharing services of
public sector bodies are excluded together with data services without a
commercial relationship between data holders and data users.\footnote{~As
  clarified in recital 28, the requirement to establish commercial
  relationships means that many technical tools for data sharing are not
  covered; these include cloud storage, analytics, data sharing software, web
  browsers and plugins for these, and email services. Instead, the goal is to
  establish data marketplaces and data ecosystems on which data is made
  available to others, as well as data pools that allow licensing for their
  use.} Also copyright-protected data is excluded.\footnote{~In other words, as
  clarified in recital 29, the regulation excludes content specified in
  Directive (EU) 2019/790. Also payment services specified in Directive (EU)
  2015/2366 are excluded together with financial instruments covered by
  Regulation (EU) 600/2014.} In general, data intermediation services are only
about sharing data; these should not use the data for purposes other than
delivery, although auxiliary functionalities such as anonymization services are
allowed~(Article~12). In other words, the goal is to promote exchange of data
via platforms, databases, and data infrastructures in general through common
protocols and data formats that ensure interoperability and security
(Article~12). The data subject rights granted by the GDPR are also specified to
apply to data intermediation services (Article~10). The establishment of these
services requires official registration with public authorities
(Article~11). The services are also supervised by competent public authorities
that the member states are obliged to designate (Article~13). Analogously to the
GDPR, according to the DGA's Article~14, these new national authorities are
empowered to impose fines and even cancel data intermediation services in case
of infringements.

The other important concept is data altruism that the member states are
instructed to promote and facilitate (Article~16). The regulation speaks about
specific data altruism organizations, which are legal persons that operate on a
not-for-profit basis without any dependencies to for-profit entities
(Article~18). As with data intermediation services, all data altruism
organizations wanting to be officially recognized as data altruism organizations
must be officially registered to public registries maintained by competent
public sector bodies (Articles 17 and 19). These organizations must keep
rigorous track of those processing data held by the organizations
(Article~20). Regarding data subjects, the regulation emphasizes that the
objectives of processing personal data are clearly defined, the geographic
location of processing is specified, and tools are provided for consent
management (Article~21). Finally, the regulation specifies compliance monitoring
requirements that are similar to those specified for data intermediation
services. That is, the member states must designate specific competent
authorities for the registration of data altruism organizations (Article~23) and
the compliance monitoring of these~(Article~24).

Coordination at the EU-level is specified to occur through the establishment of
a specific European Data Innovation Board. It is designed to operate in
cooperation with the EU-level data protection and cyber security institutions,
the new competent national public authorities, envoys of SMEs, and other related
bodies with relevant expertise, including academic institutions and civil
society groups (Article~29). A final important point is about data transfers to
countries outside of the EU. Somewhat similarly to the GDPR, such transfers
should be prevented when possible but these are still allowed based on
international agreements such as mutual legal assistance treaties
(Article~31). Given the ongoing controversies and legal cases \citep[see,
  e.g.,][]{Jurcys22}, it seems safe to assume that also the Data Governance Act
will be subjected to legal scrutiny regarding potential international data
transfers.

\section*{Reflections}

The new Data Governance Act lays down frameworks for data reuse and data
altruism under the supervision of competent public sector authorities. When
compared to other recently enacted laws, such as the Digital Services Act and
the Digital Markets Act, the legal motivation and background for the DGA are
different. In terms of EU law, perhaps the closest reference point is Directive
(EU) 2019/1024 on open data and reuse of public sector information. As noted in
recital 9 of the DGA, the member states are encouraged to follow the
directive's principle of ``open by design'' also with the new
regulation. Another related law, as noted in the DGA's recital 7, is the
Regulation (EU) 557/2013 for statistical micro-data research. Furthermore, the
new regulation resembles national laws already enacted by the member states. For
instance, Finland passed in 2019 a law for the secondary use of public sector
social and healthcare
data.\footnote{~\url{https://www.finlex.fi/fi/laki/alkup/2019/20190552}} It
shares many similarities with the DGA; the goal is to promote scientific
research and innovation, there is a specific national agency that handles the
delivery of data and its anonymization, and so forth.

The regulation further resembles the initiatives taken by non-governmental
organizations and civil society groups. Notably, the so-called
MyData\footnote{~\url{https://www.mydata.org/}} initiative, which has been
actively promoted by data and privacy activists~\citep{Lehtiniemi17,
  Lehtiniemi20}, shares similarities with the regulation and particularly its
concept of data altruism. Analogous information systems for personal data
management, data governance, and data altruism have recently been presented also
in academic research \citep{Zichichi22}. However, these initiatives, information
systems, and the DGA all seem problematic in that these rely on consent for
the sharing and processing of personal data.

Although in Europe consent builds upon the foundational concept of informational
self-determination, the use of consent as a legal basis for processing personal
data has long been criticized from empirical, legal, and technical
perspectives~\citep{Custers13, Gonzalez19, Hjerppe23SOSYM}. Good practical
examples would be the complex 20,000 word bulletproof legalese documents on one
hand and the vague click-through banners used in the current world wide web on
the other \citep{Lahtiranta18, Lundgren20}. In other words, consumers and users
of digital applications and services do not really understand to what they are
consenting to---even though the GDPR's Article 4 explicitly states that
``consent means any freely given, specific, informed and unambiguous indication
of the data subject's wishes by which he or she, by a statement or by a clear
affirmative action, signifies agreement to the processing of personal data
relating to him or her''. According to skeptical viewpoints, there is little
reason to believe that things would be different for the noble goal of data
altruism \citep{Ruohonen21MIND}. Similar points have been raised also regarding
the reuse of public sector data for which the conditions for consent are often
different~\citep{McKeown21}. In other words, it is difficult to specify the
purpose of processing personal data at the time of initial data collection in a
context that involves further processing~\citep{Mantelero15}. To some extent,
the European politicians and lawmakers seem to have been aware of these issues,
given that Article~25 in the DGA mentions the development of a specific European
data altruism consent form. However, it can be challenging for data altruism
consent to fully comply with the GDPR's consent requirements as reaching the
full potential of data economy requires flexibility in processing activities.

The DGA raises also other concerns about data protection and the GDPR. Three
such concerns deserve a brief discussion. First, the DGA seems to conflict with
some of the fundamental principles of the GDPR. In particular, the GDPR's
Article 5 explicitly states that personal data should be only collected for
``specified, explicit and legitimate purposes and not further processed in a
manner that is incompatible with those purposes''. Although the same article
specifies that this purpose limitation does not apply to public interest data
archiving, scientific research, and statistical applications, the DGA's goal of
public sector data reuse still raises a concern about whether personal data
collected by public sector bodies will be used in a manner which is unexpected
or risky to the data subjects. Given that the GDPR does not apply to anonymized
data, the DGA's provision for data reuse under the GDPR's purpose limitation
rests upon proper anonymization. The second concern follows. As is well-known,
there are efficient algorithms for de-anonymization and re-identification of
data subjects \citep{Bulmer16, Narayanan08, Rocher19}. The efficiency of such
algorithms is likely to only increase with advances in machine learning and
artificial intelligence. Hence, it remains debatable how well the
state-of-the-art privacy-preserving methods mentioned in the DGA can prevent
de-anonymization and re-identification attempts. This concern applies equally to
non-personal data held by public sector bodies under commercial
confidentiality~\citep{Kapoor21}. The last concern is about national data
protection authorities whose duties seem to substantially increase with the
DGA. For instance, according to the DGA's recital 15, prior to granting access
for reuse of data, public sector bodies should carry out data protection impact
assessments and consult data protection authorities in line with the GDPR's
Articles 35 and~36. Such consultations cover also questions about
anonymization. The DGA also mentions, in recital 26, that the new competent
bodies for monitoring intermediation services and data altruism organizations do
not have a strict supervisory function, which is reserved for data protection
authorities. Given the resourcing, coordination, and other problems already
faced by European data protection authorities~\citep{Ruohonen22IS}, a concern
remains about how well the DGA will be administrated and enforced. The problems
with the GDPR's enforcement provide an alarming precedent.

A further concern relates to the data intermediation services specified in the
regulation. Here, it remains unclear whether the existing Big Tech companies are
allowed to act as data intermediation services, and how it is possible to ensure
that such companies only provide data sharing without attempts to use the data
exchanged. Analogous concerns have already been raised in the context of
European cloud computing initiatives~\citep{Sheikh22}. Nor does the regulation
answer to a question on how SMEs and start-ups can compete against Big Tech
companies for providing data intermediation services. Furthermore, there are
potential issues related to other functionalities provided by companies acting
as data intermediation services. According to recital 33 in the DGA, a
structural separation is needed to avoid conflicts of interests; data
intermediation services should be provided through a legal person that is
separate from other activities of a given data intermediation service provider.

It remains to be seen how data altruism plays out together with the
not-for-profit data altruism organizations. The regulation is rather vague in
this regard, mainly emphasizing data protection, trust, and the idea of data
repositories for scientific and related purposes. Here, the GDPR generally acts
both as a barrier and as an enabler for data sharing and data
reuse~\citep{Vukovic22}. In general, the DGA further increases the regulatory
complexity with regard to personal data processing, particularly in the context
of scientific research. According to some surveys, Europeans have generally
positive attitudes toward reuse of their healthcare data, but have still
concerns about commercialization, security, and misuse of reused
data~\citep{Skovgaard19}. Similar results presumably apply also to voluntary
data sharing for not-for-profit purposes. Finally, there are two related
regulations in the making: the so-called Data Act and a proposal for the
creation of European health data spaces \citep{EC22b, EC22c}. The former
augments the DGA with a goal of further data sharing, portability, and
interoperability, providing also users means to gain access to data generated by
them. It may also solve some of the challenges raised in this commentary. The
second continues the same theme; the goal is to empower people by giving them
access to their health data in their home country or any other member state. In
addition, the idea is again to strengthen the single market for digital health
services and products through health data sharing across the member
states. Given the sensitivity of health data, concerns over confidentiality and
security are graver with this proposal compared to the DGA and the Data Act.

\section*{Acknowledgements}

This commentary was funded by the Strategic Research Council at the Academy of Finland (grant number 327391).

\section*{Statements and Declarations}

There is a conflict of interest with other researchers funded by the same grant (no. 327391) from the Strategic Research Council at the Academy of Finland.

\section*{Data Availability}

Not applicable.

\section*{Author Contributions}

Jukka Ruohonen wrote the initial draft. Sini Mickelsson reviewed the draft and contributed with a few insights on the legal aspects. Both authors read and approved the final manuscript.

%\bibliographystyle{sn-apacite}
%\bibliography{eu}

\begin{thebibliography}{}
\providecommand{\doi}[1]{\url{https://doi.org/#1}}
\bibcommenthead

\bibitem [\protect \citeauthoryear {%
Custers%
, {van der Hof}%
, Schermer%
, {Appleby-Arnold}%
\BCBL {}\ \BBA {} Brockdorff%
}{%
Custers%
\ \protect \BOthers {.}}{%
{\protect \APACyear {2013}}%
}]{%
Custers13}
\APACinsertmetastar {%
Custers13}%
\begin{APACrefauthors}%
Custers, B.%
, {van der Hof}, S.%
, Schermer, B.%
, {Appleby-Arnold}, S.%
\BCBL {} Brockdorff, N.%
\end{APACrefauthors}%
\unskip\
\newblock
\APACrefYearMonthDay{2013}{}{}.
\newblock
{\BBOQ}\APACrefatitle {{I}nformed {C}onsent in {S}ocial {M}edia {U}se -- {T}he
  {G}ap {B}etween {U}ser {E}xpectations and {EU} {P}ersonal {D}ata {P}rotection
  {L}aw} {{I}nformed {C}onsent in {S}ocial {M}edia {U}se -- {T}he {G}ap
  {B}etween {U}ser {E}xpectations and {EU} {P}ersonal {D}ata {P}rotection
  {L}aw}.{\BBCQ}
\newblock
\APACjournalVolNumPages{SCRIPTed}{10}{4}{435--457}.
\newblock

\newblock

\PrintBackRefs{\CurrentBib}

\bibitem [\protect \citeauthoryear {%
{European Commission}%
}{%
{European Commission}%
}{%
{\protect \APACyear {2022}}%
{\protect \APACexlab {{\protect \BCnt {1}}}}}]{%
EC22c}
\APACinsertmetastar {%
EC22c}%
\begin{APACrefauthors}%
{European Commission}%
\end{APACrefauthors}%
\unskip\
\newblock
\APACrefYearMonthDay{2022{\protect \BCnt {1}}}{}{}.
\newblock
\APACrefbtitle {{C}ommunication from the {C}ommission to the {E}uropean
  {P}arliament and the {C}ouncil: {A} {E}uropean {H}ealth {D}ata {S}pace:
  {H}arnessing the {P}ower of {H}ealth {D}ata for {P}eople, {P}atients and
  {I}nnovation.} {{C}ommunication from the {C}ommission to the {E}uropean
  {P}arliament and the {C}ouncil: {A} {E}uropean {H}ealth {D}ata {S}pace:
  {H}arnessing the {P}ower of {H}ealth {D}ata for {P}eople, {P}atients and
  {I}nnovation.}
\newblock
\APACrefnote{{COM}(2022) 196 final. {A}vailable online in January 2023:
  \url{https://health.ec.europa.eu/system/files/2022-05/com_2022-196_en.pdf}}
\PrintBackRefs{\CurrentBib}

\bibitem [\protect \citeauthoryear {%
{European Commission}%
}{%
{European Commission}%
}{%
{\protect \APACyear {2022}}%
{\protect \APACexlab {{\protect \BCnt {2}}}}}]{%
EC22b}
\APACinsertmetastar {%
EC22b}%
\begin{APACrefauthors}%
{European Commission}%
\end{APACrefauthors}%
\unskip\
\newblock
\APACrefYearMonthDay{2022{\protect \BCnt {2}}}{}{}.
\newblock
\APACrefbtitle {{P}roposal for a {R}egulation of the {E}uropean {P}arliament
  and of the {C}ouncil on {H}armonised {R}ules on {F}air {A}ccess to and {U}se
  of {D}ata ({D}ata {A}ct).} {{P}roposal for a {R}egulation of the {E}uropean
  {P}arliament and of the {C}ouncil on {H}armonised {R}ules on {F}air {A}ccess
  to and {U}se of {D}ata ({D}ata {A}ct).}
\newblock
\APACrefnote{{COM}/2022/68 final. {A}vailable online in January 2023:
  \url{https://eur-lex.europa.eu/legal-content/EN/TXT/?uri=COM\%3A2022\%3A68\%3AFIN}}
\PrintBackRefs{\CurrentBib}

\bibitem [\protect \citeauthoryear {%
{European Commission}%
}{%
{European Commission}%
}{%
{\protect \APACyear {2022}}%
{\protect \APACexlab {{\protect \BCnt {3}}}}}]{%
EC22a}
\APACinsertmetastar {%
EC22a}%
\begin{APACrefauthors}%
{European Commission}%
\end{APACrefauthors}%
\unskip\
\newblock
\APACrefYearMonthDay{2022{\protect \BCnt {3}}}{}{}.
\newblock
\APACrefbtitle {{R}egulation {(EU)} 2022/868 of the {E}uropean {P}arliament and
  of the {C}ouncil of 30 {M}ay 2022 on {E}uropean {D}ata {G}overnance and
  {A}mending {R}egulation {(EU)} 2018/1724 ({D}ata {G}overnance {A}ct).}
  {{R}egulation {(EU)} 2022/868 of the {E}uropean {P}arliament and of the
  {C}ouncil of 30 {M}ay 2022 on {E}uropean {D}ata {G}overnance and {A}mending
  {R}egulation {(EU)} 2018/1724 ({D}ata {G}overnance {A}ct).}
\newblock
\APACrefnote{{A}vailable online in January 2023:
  \url{https://eur-lex.europa.eu/legal-content/EN/TXT/?uri=CELEX\%3A32022R0868&qid=1673517089264}}
\PrintBackRefs{\CurrentBib}

\bibitem [\protect \citeauthoryear {%
Gonz\'alez%
\ \BBA {} {de Hert}%
}{%
Gonz\'alez%
\ \BBA {} {de Hert}%
}{%
{\protect \APACyear {2019}}%
}]{%
Gonzalez19}
\APACinsertmetastar {%
Gonzalez19}%
\begin{APACrefauthors}%
Gonz\'alez, E.G.%
\BCBT {}\ \BBA {} {de Hert}, P.%
\end{APACrefauthors}%
\unskip\
\newblock
\APACrefYearMonthDay{2019}{}{}.
\newblock
{\BBOQ}\APACrefatitle {{U}nderstanding the {L}egal {P}rovisions that {A}llow
  {P}rocessing and {P}rofiling of {P}ersonal {D}ata---{A}n {A}nalysis of {GDPR}
  {P}rovisions and {P}rinciples} {{U}nderstanding the {L}egal {P}rovisions that
  {A}llow {P}rocessing and {P}rofiling of {P}ersonal {D}ata---{A}n {A}nalysis
  of {GDPR} {P}rovisions and {P}rinciples}.{\BBCQ}
\newblock
\APACjournalVolNumPages{ERA Forum}{19}{}{597--621}.
\newblock

\newblock

\PrintBackRefs{\CurrentBib}

\bibitem [\protect \citeauthoryear {%
{Henriksen-Bulmer}%
\ \BBA {} Jeary%
}{%
{Henriksen-Bulmer}%
\ \BBA {} Jeary%
}{%
{\protect \APACyear {2016}}%
}]{%
Bulmer16}
\APACinsertmetastar {%
Bulmer16}%
\begin{APACrefauthors}%
{Henriksen-Bulmer}, J.%
\BCBT {}\ \BBA {} Jeary, S.%
\end{APACrefauthors}%
\unskip\
\newblock
\APACrefYearMonthDay{2016}{}{}.
\newblock
{\BBOQ}\APACrefatitle {{R}e-{I}dentification {A}ttacks---{A} {S}ystematic
  {L}iterature {R}eview} {{R}e-{I}dentification {A}ttacks---{A} {S}ystematic
  {L}iterature {R}eview}.{\BBCQ}
\newblock
\APACjournalVolNumPages{International Journal of Information
  Management}{36}{6}{1184--1192}.
\newblock

\newblock

\PrintBackRefs{\CurrentBib}

\bibitem [\protect \citeauthoryear {%
Hjerppe%
, Ruohonen%
\BCBL {}\ \BBA {} Lepp\"anen%
}{%
Hjerppe%
\ \protect \BOthers {.}}{%
{\protect \APACyear {2023}}%
}]{%
Hjerppe23SOSYM}
\APACinsertmetastar {%
Hjerppe23SOSYM}%
\begin{APACrefauthors}%
Hjerppe, K.%
, Ruohonen, J.%
\BCBL {} Lepp\"anen, V.%
\end{APACrefauthors}%
\unskip\
\newblock
\APACrefYearMonthDay{2023}{}{}.
\newblock
{\BBOQ}\APACrefatitle {{E}xtracting {LPL} {P}rivacy {P}olicy {P}urposes from
  {A}nnotated {W}eb {S}ervice {S}ource {C}ode} {{E}xtracting {LPL} {P}rivacy
  {P}olicy {P}urposes from {A}nnotated {W}eb {S}ervice {S}ource {C}ode}.{\BBCQ}
\newblock
\APACjournalVolNumPages{Software and Systems Modeling}{22}{}{331--349}.
\newblock

\newblock

\PrintBackRefs{\CurrentBib}

\bibitem [\protect \citeauthoryear {%
Jurcys%
, Compagnucci%
\BCBL {}\ \BBA {} Fenwick%
}{%
Jurcys%
\ \protect \BOthers {.}}{%
{\protect \APACyear {2022}}%
}]{%
Jurcys22}
\APACinsertmetastar {%
Jurcys22}%
\begin{APACrefauthors}%
Jurcys, P.%
, Compagnucci, M.C.%
\BCBL {} Fenwick, M.%
\end{APACrefauthors}%
\unskip\
\newblock
\APACrefYearMonthDay{2022}{}{}.
\newblock
{\BBOQ}\APACrefatitle {{T}he {F}uture of {I}nternational {D}ata {T}ransfers:
  {M}anaging {L}egal {R}isk with a '{U}ser-{H}eld' {D}ata {M}odel} {{T}he
  {F}uture of {I}nternational {D}ata {T}ransfers: {M}anaging {L}egal {R}isk
  with a '{U}ser-{H}eld' {D}ata {M}odel}.{\BBCQ}
\newblock
\APACjournalVolNumPages{Computer Law \& Security Review}{46}{}{105691}.
\newblock

\newblock

\PrintBackRefs{\CurrentBib}

\bibitem [\protect \citeauthoryear {%
Kapoor%
\ \BBA {} Nanda%
}{%
Kapoor%
\ \BBA {} Nanda%
}{%
{\protect \APACyear {2021}}%
}]{%
Kapoor21}
\APACinsertmetastar {%
Kapoor21}%
\begin{APACrefauthors}%
Kapoor, A.%
\BCBT {}\ \BBA {} Nanda, A.%
\end{APACrefauthors}%
\unskip\
\newblock
\APACrefYearMonthDay{2021}{}{}.
\newblock
{\BBOQ}\APACrefatitle {{N}on-{P}ersonal {D}ata {S}haring: {P}otential,
  {P}athways and {P}roblems} {{N}on-{P}ersonal {D}ata {S}haring: {P}otential,
  {P}athways and {P}roblems}.{\BBCQ}
\newblock
\APACjournalVolNumPages{CSI Transactions on ICT}{9}{}{165--169}.
\newblock

\newblock

\PrintBackRefs{\CurrentBib}

\bibitem [\protect \citeauthoryear {%
K\"onig%
}{%
K\"onig%
}{%
{\protect \APACyear {2022}}%
}]{%
Konig22}
\APACinsertmetastar {%
Konig22}%
\begin{APACrefauthors}%
K\"onig, P.D.%
\end{APACrefauthors}%
\unskip\
\newblock
\APACrefYearMonthDay{2022}{}{}.
\newblock
{\BBOQ}\APACrefatitle {{F}ortress {E}urope 4.0? {A}n {A}nalysis of {EU} {D}ata
  {G}overnance {T}hrough the {L}ens of the {R}esource {R}egime {C}oncept}
  {{F}ortress {E}urope 4.0? {A}n {A}nalysis of {EU} {D}ata {G}overnance
  {T}hrough the {L}ens of the {R}esource {R}egime {C}oncept}.{\BBCQ}
\newblock
\APACjournalVolNumPages{European Policy Analysis}{8}{4}{484--504}.
\newblock

\newblock

\PrintBackRefs{\CurrentBib}

\bibitem [\protect \citeauthoryear {%
Lahtiranta%
\ \BBA {} Hyrynsalmi%
}{%
Lahtiranta%
\ \BBA {} Hyrynsalmi%
}{%
{\protect \APACyear {2018}}%
}]{%
Lahtiranta18}
\APACinsertmetastar {%
Lahtiranta18}%
\begin{APACrefauthors}%
Lahtiranta, J.%
\BCBT {}\ \BBA {} Hyrynsalmi, S.%
\end{APACrefauthors}%
\unskip\
\newblock
\APACrefYearMonthDay{2018}{}{}.
\newblock
{\BBOQ}\APACrefatitle {{C}rude and {R}ude?} {{C}rude and {R}ude?}{\BBCQ}
\newblock
\APACjournalVolNumPages{Communications of the ACM}{61}{11}{34--35}.
\newblock

\newblock

\PrintBackRefs{\CurrentBib}

\bibitem [\protect \citeauthoryear {%
Lambach%
\ \BBA {} Oppermann%
}{%
Lambach%
\ \BBA {} Oppermann%
}{%
{\protect \APACyear {2022}}%
}]{%
Lambach22}
\APACinsertmetastar {%
Lambach22}%
\begin{APACrefauthors}%
Lambach, D.%
\BCBT {}\ \BBA {} Oppermann, K.%
\end{APACrefauthors}%
\unskip\
\newblock
\APACrefYearMonthDay{2022}{}{}.
\newblock
{\BBOQ}\APACrefatitle {{N}arratives of {D}igital {S}overeignty in {G}erman
  {P}olitical {D}iscourse} {{N}arratives of {D}igital {S}overeignty in {G}erman
  {P}olitical {D}iscourse}.{\BBCQ}
\newblock
\APACjournalVolNumPages{Governance}{}{Published online in April}{1--17}.
\newblock

\newblock

\PrintBackRefs{\CurrentBib}

\bibitem [\protect \citeauthoryear {%
Lehtiniemi%
}{%
Lehtiniemi%
}{%
{\protect \APACyear {2017}}%
}]{%
Lehtiniemi17}
\APACinsertmetastar {%
Lehtiniemi17}%
\begin{APACrefauthors}%
Lehtiniemi, T.%
\end{APACrefauthors}%
\unskip\
\newblock
\APACrefYearMonthDay{2017}{}{}.
\newblock
{\BBOQ}\APACrefatitle {{P}ersonal {D}ata {S}paces: {A}n {I}ntervention in
  {S}urveillance {C}apitalism?} {{P}ersonal {D}ata {S}paces: {A}n
  {I}ntervention in {S}urveillance {C}apitalism?}{\BBCQ}
\newblock
\APACjournalVolNumPages{Surveillance \& Society}{15}{17}{626--639}.
\newblock

\newblock

\PrintBackRefs{\CurrentBib}

\bibitem [\protect \citeauthoryear {%
Lehtiniemi%
\ \BBA {} Haapoja%
}{%
Lehtiniemi%
\ \BBA {} Haapoja%
}{%
{\protect \APACyear {2020}}%
}]{%
Lehtiniemi20}
\APACinsertmetastar {%
Lehtiniemi20}%
\begin{APACrefauthors}%
Lehtiniemi, T.%
\BCBT {}\ \BBA {} Haapoja, J.%
\end{APACrefauthors}%
\unskip\
\newblock
\APACrefYearMonthDay{2020}{}{}.
\newblock
{\BBOQ}\APACrefatitle {{D}ata {A}gency at {S}take: {M}y{D}ata {A}ctivism and
  {A}lternative {F}rames of {E}qual {P}articipation} {{D}ata {A}gency at
  {S}take: {M}y{D}ata {A}ctivism and {A}lternative {F}rames of {E}qual
  {P}articipation}.{\BBCQ}
\newblock
\APACjournalVolNumPages{New Media \& Society}{22}{1}{87--104}.
\newblock

\newblock

\PrintBackRefs{\CurrentBib}

\bibitem [\protect \citeauthoryear {%
Lundgren%
}{%
Lundgren%
}{%
{\protect \APACyear {2020}}%
}]{%
Lundgren20}
\APACinsertmetastar {%
Lundgren20}%
\begin{APACrefauthors}%
Lundgren, B.%
\end{APACrefauthors}%
\unskip\
\newblock
\APACrefYearMonthDay{2020}{}{}.
\newblock
{\BBOQ}\APACrefatitle {{H}ow {S}oftware {D}evelopers {C}an {F}ix {P}art of
  {GDPR}'s {P}roblem of {C}lick-{T}hrough {C}onsents} {{H}ow {S}oftware
  {D}evelopers {C}an {F}ix {P}art of {GDPR}'s {P}roblem of {C}lick-{T}hrough
  {C}onsents}.{\BBCQ}
\newblock
\APACjournalVolNumPages{AI \& SOCIETY}{35}{}{759--760}.
\newblock

\newblock

\PrintBackRefs{\CurrentBib}

\bibitem [\protect \citeauthoryear {%
Mantelero%
\ \BBA {} Vaciago%
}{%
Mantelero%
\ \BBA {} Vaciago%
}{%
{\protect \APACyear {2015}}%
}]{%
Mantelero15}
\APACinsertmetastar {%
Mantelero15}%
\begin{APACrefauthors}%
Mantelero, A.%
\BCBT {}\ \BBA {} Vaciago, G.%
\end{APACrefauthors}%
\unskip\
\newblock
\APACrefYearMonthDay{2015}{}{}.
\newblock
{\BBOQ}\APACrefatitle {{D}ata {P}rotection in a {B}ig {D}ata {S}ociety. {I}deas
  for a {F}uture {R}egulation} {{D}ata {P}rotection in a {B}ig {D}ata
  {S}ociety. {I}deas for a {F}uture {R}egulation}.{\BBCQ}
\newblock
\APACjournalVolNumPages{Digital Investigation}{15}{}{104--109}.
\newblock

\newblock

\PrintBackRefs{\CurrentBib}

\bibitem [\protect \citeauthoryear {%
McKeown%
\ \protect \BOthers {.}}{%
McKeown%
\ \protect \BOthers {.}}{%
{\protect \APACyear {2021}}%
}]{%
McKeown21}
\APACinsertmetastar {%
McKeown21}%
\begin{APACrefauthors}%
McKeown, A.%
, Mourby, M.%
, Harrison, P.%
, Walker, S.%
, Sheehan, M.%
\BCBL {} Singh, I.%
\end{APACrefauthors}%
\unskip\
\newblock
\APACrefYearMonthDay{2021}{}{}.
\newblock
{\BBOQ}\APACrefatitle {{E}thical {I}ssues in {C}onsent for the {R}euse of
  {D}ata in {H}ealth {D}ata {P}latforms} {{E}thical {I}ssues in {C}onsent for
  the {R}euse of {D}ata in {H}ealth {D}ata {P}latforms}.{\BBCQ}
\newblock
\APACjournalVolNumPages{Science and Engineering Ethics}{27}{}{1--21}.
\newblock

\newblock

\PrintBackRefs{\CurrentBib}

\bibitem [\protect \citeauthoryear {%
Narayanan%
\ \BBA {} Shmatikov%
}{%
Narayanan%
\ \BBA {} Shmatikov%
}{%
{\protect \APACyear {2008}}%
}]{%
Narayanan08}
\APACinsertmetastar {%
Narayanan08}%
\begin{APACrefauthors}%
Narayanan, A.%
\BCBT {}\ \BBA {} Shmatikov, V.%
\end{APACrefauthors}%
\unskip\
\newblock
\APACrefYearMonthDay{2008}{}{}.
\newblock
{\BBOQ}\APACrefatitle {{R}obust {D}e-{A}nonymization of {L}arge {S}parse
  {D}atasets} {{R}obust {D}e-{A}nonymization of {L}arge {S}parse
  {D}atasets}.{\BBCQ}
\newblock
 \APACrefbtitle {{P}roceedings of the {IEEE} {S}ymposium on {S}ecurity and
  {P}rivacy ({S\&P} 2008)} {{P}roceedings of the {IEEE} {S}ymposium on
  {S}ecurity and {P}rivacy ({S\&P} 2008)}\ (\BPGS\ 111--125).
\newblock
\APACaddressPublisher{Oakland}{IEEE}.
\PrintBackRefs{\CurrentBib}

\bibitem [\protect \citeauthoryear {%
Rocher%
, Hendrickx%
\BCBL {}\ \BBA {} {de Montjoye}%
}{%
Rocher%
\ \protect \BOthers {.}}{%
{\protect \APACyear {2019}}%
}]{%
Rocher19}
\APACinsertmetastar {%
Rocher19}%
\begin{APACrefauthors}%
Rocher, L.%
, Hendrickx, J.M.%
\BCBL {} {de Montjoye}, Y.%
\end{APACrefauthors}%
\unskip\
\newblock
\APACrefYearMonthDay{2019}{}{}.
\newblock
{\BBOQ}\APACrefatitle {{E}stimating the {S}uccess of {R}e-{I}dentifications in
  {I}ncomplete {D}atasets {U}sing {G}enerative {M}odels} {{E}stimating the
  {S}uccess of {R}e-{I}dentifications in {I}ncomplete {D}atasets {U}sing
  {G}enerative {M}odels}.{\BBCQ}
\newblock
\APACjournalVolNumPages{Nature Communications}{10}{}{3069}.
\newblock

\newblock

\PrintBackRefs{\CurrentBib}

\bibitem [\protect \citeauthoryear {%
Ruohonen%
}{%
Ruohonen%
}{%
{\protect \APACyear {2021}}%
}]{%
Ruohonen21MIND}
\APACinsertmetastar {%
Ruohonen21MIND}%
\begin{APACrefauthors}%
Ruohonen, J.%
\end{APACrefauthors}%
\unskip\
\newblock
\APACrefYearMonthDay{2021}{}{}.
\newblock
{\BBOQ}\APACrefatitle {{T}he {T}reatchery of {I}mages in the {D}igital
  {S}overeignty {D}ebate} {{T}he {T}reatchery of {I}mages in the {D}igital
  {S}overeignty {D}ebate}.{\BBCQ}
\newblock
\APACjournalVolNumPages{Minds and Machines}{31}{}{439--456}.
\newblock

\newblock

\PrintBackRefs{\CurrentBib}

\bibitem [\protect \citeauthoryear {%
Ruohonen%
\ \BBA {} Hjerppe%
}{%
Ruohonen%
\ \BBA {} Hjerppe%
}{%
{\protect \APACyear {2022}}%
}]{%
Ruohonen22IS}
\APACinsertmetastar {%
Ruohonen22IS}%
\begin{APACrefauthors}%
Ruohonen, J.%
\BCBT {}\ \BBA {} Hjerppe, K.%
\end{APACrefauthors}%
\unskip\
\newblock
\APACrefYearMonthDay{2022}{}{}.
\newblock
{\BBOQ}\APACrefatitle {{T}he {GDPR} {E}nforcement {F}ines at {G}lance} {{T}he
  {GDPR} {E}nforcement {F}ines at {G}lance}.{\BBCQ}
\newblock
\APACjournalVolNumPages{Information Systems}{106}{}{101876}.
\newblock

\newblock

\PrintBackRefs{\CurrentBib}

\bibitem [\protect \citeauthoryear {%
Sheikh%
}{%
Sheikh%
}{%
{\protect \APACyear {2022}}%
}]{%
Sheikh22}
\APACinsertmetastar {%
Sheikh22}%
\begin{APACrefauthors}%
Sheikh, H.%
\end{APACrefauthors}%
\unskip\
\newblock
\APACrefYearMonthDay{2022}{}{}.
\newblock
{\BBOQ}\APACrefatitle {{E}uropean {D}igital {S}overeignty: {A} {L}ayered
  {A}pproach} {{E}uropean {D}igital {S}overeignty: {A} {L}ayered
  {A}pproach}.{\BBCQ}
\newblock
\APACjournalVolNumPages{Digital Society}{1}{3}{1--25}.
\newblock

\newblock

\PrintBackRefs{\CurrentBib}

\bibitem [\protect \citeauthoryear {%
Skovgaard%
, Wadmann%
\BCBL {}\ \BBA {} Hoeyer%
}{%
Skovgaard%
\ \protect \BOthers {.}}{%
{\protect \APACyear {2019}}%
}]{%
Skovgaard19}
\APACinsertmetastar {%
Skovgaard19}%
\begin{APACrefauthors}%
Skovgaard, L.L.%
, Wadmann, S.%
\BCBL {} Hoeyer, K.%
\end{APACrefauthors}%
\unskip\
\newblock
\APACrefYearMonthDay{2019}{}{}.
\newblock
{\BBOQ}\APACrefatitle {{A} {R}eview of {A}ttitudes {T}owards the {R}euse of
  {H}ealth {D}ata {A}mong {P}eople in the {E}uropean {U}nion: {T}he {P}rimacy
  of {P}urpose and the {C}ommon {G}ood} {{A} {R}eview of {A}ttitudes {T}owards
  the {R}euse of {H}ealth {D}ata {A}mong {P}eople in the {E}uropean {U}nion:
  {T}he {P}rimacy of {P}urpose and the {C}ommon {G}ood}.{\BBCQ}
\newblock
\APACjournalVolNumPages{Health Policy}{123}{6}{564--571}.
\newblock

\newblock

\PrintBackRefs{\CurrentBib}

\bibitem [\protect \citeauthoryear {%
Starkbaum%
\ \BBA {} Felt%
}{%
Starkbaum%
\ \BBA {} Felt%
}{%
{\protect \APACyear {2019}}%
}]{%
Starkbaum19}
\APACinsertmetastar {%
Starkbaum19}%
\begin{APACrefauthors}%
Starkbaum, J.%
\BCBT {}\ \BBA {} Felt, U.%
\end{APACrefauthors}%
\unskip\
\newblock
\APACrefYearMonthDay{2019}{}{}.
\newblock
{\BBOQ}\APACrefatitle {{N}egotiating the {R}euse of {H}ealth-{D}ata:
  {R}esearch, {B}ig {D}ata, and the {E}uropean {G}eneral {D}ata {P}rotection
  {R}egulation} {{N}egotiating the {R}euse of {H}ealth-{D}ata: {R}esearch,
  {B}ig {D}ata, and the {E}uropean {G}eneral {D}ata {P}rotection
  {R}egulation}.{\BBCQ}
\newblock
\APACjournalVolNumPages{Big Data \& Society}{6}{2}{2053951719862594}.
\newblock

\newblock

\PrintBackRefs{\CurrentBib}

\bibitem [\protect \citeauthoryear {%
{van Ooijen}%
\ \BBA {} Vrabec%
}{%
{van Ooijen}%
\ \BBA {} Vrabec%
}{%
{\protect \APACyear {2019}}%
}]{%
vanOoijen19}
\APACinsertmetastar {%
vanOoijen19}%
\begin{APACrefauthors}%
{van Ooijen}, I.%
\BCBT {}\ \BBA {} Vrabec, H.U.%
\end{APACrefauthors}%
\unskip\
\newblock
\APACrefYearMonthDay{2019}{}{}.
\newblock
{\BBOQ}\APACrefatitle {{D}oes the {GDPR} {E}nhance {C}onsumers' {C}ontrol over
  {P}ersonal {D}ata? {A}n {A}nalysis from a {B}ehavioural {P}erspective}
  {{D}oes the {GDPR} {E}nhance {C}onsumers' {C}ontrol over {P}ersonal {D}ata?
  {A}n {A}nalysis from a {B}ehavioural {P}erspective}.{\BBCQ}
\newblock
\APACjournalVolNumPages{Journal of Consumer Policy}{42}{}{91--107}.
\newblock

\newblock

\PrintBackRefs{\CurrentBib}

\bibitem [\protect \citeauthoryear {%
Vukovic%
, Ivankovic%
, Habl%
\BCBL {}\ \BBA {} Dimnjakovic%
}{%
Vukovic%
\ \protect \BOthers {.}}{%
{\protect \APACyear {2022}}%
}]{%
Vukovic22}
\APACinsertmetastar {%
Vukovic22}%
\begin{APACrefauthors}%
Vukovic, J.%
, Ivankovic, D.%
, Habl, C.%
\BCBL {} Dimnjakovic, J.%
\end{APACrefauthors}%
\unskip\
\newblock
\APACrefYearMonthDay{2022}{}{}.
\newblock
{\BBOQ}\APACrefatitle {{E}nablers and {B}arriers to the {S}econdary {U}se of
  {H}ealth {D}ata in {E}urope: {G}eneral {D}ata {P}rotection {R}egulation
  {P}erspective} {{E}nablers and {B}arriers to the {S}econdary {U}se of
  {H}ealth {D}ata in {E}urope: {G}eneral {D}ata {P}rotection {R}egulation
  {P}erspective}.{\BBCQ}
\newblock
\APACjournalVolNumPages{Archives of Public Health}{80}{}{115}.
\newblock

\newblock

\PrintBackRefs{\CurrentBib}

\bibitem [\protect \citeauthoryear {%
Zichichi%
, Ferretti%
, {D'Angelo}%
\BCBL {}\ \BBA {} {Rodr\'iguez-Doncel}%
}{%
Zichichi%
\ \protect \BOthers {.}}{%
{\protect \APACyear {2022}}%
}]{%
Zichichi22}
\APACinsertmetastar {%
Zichichi22}%
\begin{APACrefauthors}%
Zichichi, M.%
, Ferretti, S.%
, {D'Angelo}, G.%
\BCBL {} {Rodr\'iguez-Doncel}, V.%
\end{APACrefauthors}%
\unskip\
\newblock
\APACrefYearMonthDay{2022}{}{}.
\newblock
{\BBOQ}\APACrefatitle {{D}ata {G}overnance {T}hrough a {M}ulti-{DLT}
  {A}rchitecture in {V}iew of the {GDPR}} {{D}ata {G}overnance {T}hrough a
  {M}ulti-{DLT} {A}rchitecture in {V}iew of the {GDPR}}.{\BBCQ}
\newblock
\APACjournalVolNumPages{Cluster Computing}{25}{}{4515--4542}.
\newblock

\newblock

\PrintBackRefs{\CurrentBib}

\end{thebibliography}

\end{document}